\def\DATA{\number\day\ \ifcase\month\or{}gennaio{}\or%
febbraio{}\or{}marzo{}\or{}aprile\or{}maggio{}\or{}giugno{}\or{}luglio{}\or%
agosto{}\or{}settembre{}\or{}ottobre{}\or{}novembre{}\or{}dicembre\fi%
\ \number\year}
\def\Di{\DATA}
\newcount\mgnf\newcount\tipi\newcount\tipoformule
\newcount\aux\newcount\piepagina\newcount\xdata
\mgnf=1
\aux=1           
\tipoformule=1   
\piepagina=2     
\xdata=0         

\ifnum\mgnf=1 \aux=0 \tipoformule =1 \piepagina=2 \xdata=1\fi
\newcount\bibl
\ifnum\mgnf=0\bibl=0\else\bibl=1\fi
\bibl=0

%
%
%
%
\ifnum\bibl=0
\def\ref#1#2#3{[#1#2]\write8{#1@#2}}
\def\rif#1#2#3#4{\item{[#1#2]} #3}
\fi

\ifnum\bibl=1
\def\ref#1#2#3{[#3]\write8{#1@#2}}
\def\rif#1#2#3#4{}

\fi

\def\9#1{\ifnum\aux=1#1\else\relax\fi}
\ifnum\piepagina=0 \footline={\rlap{\hbox{\copy200}\
$\st[\number\pageno]$}\hss\tenrm \foglio\hss}\fi \ifnum\piepagina=1
\footline={\rlap{\hbox{\copy200}} \hss\tenrm \folio\hss}\fi
\ifnum\piepagina=2\footline{\hss\tenrm\folio\hss}\fi

\ifnum\mgnf=0 \magnification=\magstep0
\hsize=13.5truecm\vsize=22.5truecm \parindent=4.pt\fi
\ifnum\mgnf=1 \magnification=\magstep1
\hsize=16.0truecm\vsize=22.5truecm\baselineskip14pt
\parindent=4.pt\fi

\let\a=\alpha\let\b=\beta \let\g=\gamma \let\d=\delta
\let\e=\varepsilon \let\z=\zeta \let\h=\eta
\let\th=\vartheta\let\k=\kappa \let\l=\lambda \let\m=\mu 
 \let\p=\pi \let\r=\rho \let\s=\sigma 
   
  \let\D=\Delta 
\let\L=\Lambda   
  
{\count255=\time\divide\count255 by 60 \xdef\oramin{\number\count255}
\multiply\count255 by-60\advance\count255 by\time
\xdef\oramin{\oramin:\ifnum\count255<10 0\fi\the\count255}}
\def\ora{\oramin }

\ifnum\xdata=0
\def\data{\number\day/\ifcase\month\or gennaio \or
febbraio \or marzo \or aprile \or maggio \or giugno \or luglio \or
agosto \or settembre \or ottobre \or novembre \or dicembre
\fi/\number\year;\ \ora}
\else
\def\data{\Di}
\fi

\setbox200\hbox{$\scriptscriptstyle \data $}
\newcount\pgn \pgn=1
\def\foglio{\number\numsec:\number\pgn
\global\advance\pgn by 1} \def\foglioa{A\number\numsec:\number\pgn
\global\advance\pgn by 1}
\global\newcount\numsec\global\newcount\numfor \global\newcount\numfig
\gdef\profonditastruttura{\dp\strutbox}
\def\senondefinito#1{\expandafter\ifx\csname#1\endcsname\relax}
\def\SIA #1,#2,#3 {\senondefinito{#1#2} \expandafter\xdef\csname
#1#2\endcsname{#3} \else \write16{???? ma #1,#2 e' gia' stato definito
!!!!} \fi} \def\etichetta(#1){(\veroparagrafo.\veraformula) \SIA
e,#1,(\veroparagrafo.\veraformula) \global\advance\numfor by 1
\9{\write15{\string\FU (#1){\equ(#1)}}} \9{ \write16{ EQ \equ(#1) == #1
}}} \def \FU(#1)#2{\SIA fu,#1,#2 }
\def\etichettaa(#1){(A\veroparagrafo.\veraformula) \SIA
e,#1,(A\veroparagrafo.\veraformula) \global\advance\numfor by 1
\9{\write15{\string\FU (#1){\equ(#1)}}} \9{ \write16{ EQ \equ(#1) == #1
}}} \def\getichetta(#1){Fig.  \verafigura \SIA e,#1,{\verafigura}
\global\advance\numfig by 1 \9{\write15{\string\FU (#1){\equ(#1)}}} \9{
\write16{ Fig.  \equ(#1) ha simbolo #1 }}} \newdimen\gwidth \def\BOZZA{
\def\alato(##1){ {\vtop to \profonditastruttura{\baselineskip
\profonditastruttura\vss
\rlap{\kern-\hsize\kern-1.2truecm{$\scriptstyle##1$}}}}}
\def\galato(##1){ \gwidth=\hsize \divide\gwidth by 2 {\vtop to
\profonditastruttura{\baselineskip \profonditastruttura\vss
\rlap{\kern-\gwidth\kern-1.2truecm{$\scriptstyle##1$}}}}} }
\def\alato(#1){} \def\galato(#1){}
\def\veroparagrafo{\number\numsec}\def\veraformula{\number\numfor}
\def\verafigura{\number\numfig}
\def\geq(#1){\getichetta(#1)\galato(#1)}
\def\Eq(#1){\eqno{\etichetta(#1)\alato(#1)}}
\def\eq(#1){\etichetta(#1)\alato(#1)}
\def\Eqa(#1){\eqno{\etichettaa(#1)\alato(#1)}}
\def\eqa(#1){\etichettaa(#1)\alato(#1)}
\def\eqv(#1){\senondefinito{fu#1}$\clubsuit$#1\write16{No translation
for #1} \else\csname fu#1\endcsname\fi}
\def\equ(#1){\senondefinito{e#1}\eqv(#1)\else\csname e#1\endcsname\fi}
\openin13=#1.aux \ifeof13 \relax \else \input #1.aux \closein13\fi
\openin14=\jobname.aux \ifeof14 \relax \else \input \jobname.aux
\closein14 \fi \9{\openout15=\jobname.aux} \newskip\ttglue


\ifnum\mgnf=1\font\titolone=cmbx12 scaled \magstep3
\font\titolo=cmbxti10 scaled \magstep1\fi
\font\ottorm=cmr8\font\ottoi=cmmi7\font\ottosy=cmsy7
\font\ottobf=cmbx7\font\ottott=cmtt8\font\ottosl=cmsl8\font\ottoit=cmti7
\font\sixrm=cmr6\font\sixbf=cmbx7\font\sixi=cmmi7\font\sixsy=cmsy7

\font\fiverm=cmr5\font\fivesy=cmsy5\font\fivei=cmmi5\font\fivebf=cmbx5
\def\ottopunti{\def\rm{\fam0\ottorm}\textfont0=\ottorm%
\scriptfont0=\sixrm\scriptscriptfont0=\fiverm\textfont1=\ottoi%
\scriptfont1=\sixi\scriptscriptfont1=\fivei\textfont2=\ottosy%
\scriptfont2=\sixsy\scriptscriptfont2=\fivesy\textfont3=\tenex%
\scriptfont3=\tenex\scriptscriptfont3=\tenex\textfont\itfam=\ottoit%
\def\it{\fam\itfam\ottoit}\textfont\slfam=\ottosl%
\def\sl{\fam\slfam\ottosl}\textfont\ttfam=\ottott%
\def\tt{\fam\ttfam\ottott}\textfont\bffam=\ottobf%
\scriptfont\bffam=\sixbf\scriptscriptfont\bffam=\fivebf%
\def\bf{\fam\bffam\ottobf}\tt\ttglue=.5em plus.25em minus.15em%
\setbox\strutbox=\hbox{\vrule height7pt depth2pt width0pt}%
\normalbaselineskip=9pt\let\sc=\sixrm\normalbaselines\rm}
\font\cs=cmcsc10

\catcode`@=11
\def\footnote#1{\edef\@sf{\spacefactor\the\spacefactor}#1\@sf
\insert\footins\bgroup\ottopunti\interlinepenalty100\let\par=\endgraf
\leftskip=0pt \rightskip=0pt \splittopskip=10pt plus 1pt minus 1pt
\floatingpenalty=20000
\smallskip\item{#1}\bgroup\strut\aftergroup\@foot\let\next}
\skip\footins=12pt plus 2pt minus 4pt\dimen\footins=30pc\catcode`@=12

\newdimen\xshift \newdimen\xwidth \newdimen\yshift
\def\ins#1#2#3{\vbox to0pt{\kern-#2pt \hbox{\kern#1pt #3}\vss}\nointerlineskip}

\def\eqfig#1#2#3#4#5{ \par\xwidth=#1pt
\xshift=\hsize \advance\xshift by-\xwidth \divide\xshift by 2
\yshift=#2pt \divide\yshift by 2 \line{\hglue\xshift \vbox to #2pt{\vfil #3
\includegraphics{#4} }\hfill\raise\yshift\hbox{#5}}}

\newwrite\figura   \def\8{\immediate\write\figura\bgroup}


\def\V#1{{\,\underline#1\,}}
\def\T#1{#1\kern-4pt\lower9pt\hbox{$\widetilde{}$}\kern4pt{}}
\let\dpr=\partial \let\io=\infty
\def\fra#1#2{{#1\over#2}}\def\media#1{\langle{#1}\rangle}\let\0=\noindent
\def\guida{\leaders\hbox to 1em{\hss.\hss}\hfill}
\def\tende#1{\vtop{\ialign{##\crcr\rightarrowfill\crcr
\noalign{\kern-1pt\nointerlineskip} \hglue3.pt${\scriptstyle
#1}$\hglue3.pt\crcr}}} \def\otto{\
{\kern-1.truept\leftarrow\kern-5.truept\to\kern-1.truept}\ }

\def\tto{{\Rightarrow}}
\def\pagina{\vfill\eject}\let\ciao=\bye

\def\st{\scriptscriptstyle}
\def\*{\vskip0.3truecm}

\def\lis#1{{\overline #1}}\def\eg{\hbox{\it e.g.\ }}

\def\ie{\hbox{\it i.e.\ }}

\def\fiat{{}}
\def\\{\hfill\break} \def\={{ \; \equiv \; }}

\def\annota#1{\footnote{${{}^{{\bf#1\relax \rm}}}$}}
\ifnum\aux=1\BOZZA\else\relax\fi
\ifnum\tipoformule=1\let\Eq=\eqno\def\eq{}\let\Eqa=\eqno\def\eqa{}
\def\equ{{}}\fi
\def\defi{\,{\buildrel def \over =}\,}

\def\1{\ifnum\mgnf=0\pagina\else\relax\fi}
\def\W#1{#1_{\kern-3pt\lower6.6truept\hbox to 1.1truemm
{$\widetilde{}$\hfill}}\kern2pt\,}

\def\xx{{\V x}}

\def\NN{{\cal N}}

\def\aa{{\V \a}}

\def\ndpr{{\kern1pt\raise 1pt\hbox{$\not$}\kern1pt\dpr\kern1pt}}
\def\Ndpr{{\kern1pt\raise 1pt\hbox{$\not$}\kern0.3pt\dpr\kern1pt}}

\def\LL{{\cal L}}

\def\xx{{\V x}}\def\PP{{\cal P}}\def\yy{{\V y}}
\fiat

\centerline{\titolone A local f\/luctuation theorem}
\*\*

\centerline{\titolo G. Gallavotti}
\centerline{\it Fisica, Universit\`a di Roma 1}
\centerline{\Di}
\vskip.8truecm
\line{\vtop{
\line{\hskip1.5truecm\vbox{\advance \hsize by -3.1 truecm
\\{\cs Abstract.}
{\it A mechanism for the validity of a local version of the
fluctuation theorem, uniform in the system size, is discussed for a
reversible chain of weakly coupled Anosov systems.}}\hfill}}}
\vskip1.5truecm

\0{\bf 1. Introduction.}
\numsec=1\numfor=1\*

In recent papers, [GC], [MR], a general law governing the f\/luctuations
of phase space volume contractions in a dynamical system (\ie a smooth
smoothly invertible map $S: \,x\to Sx$ on a phase space $M$) has been
derived under various {\it chaoticity} assumptions. Its generality makes
it remarkable. Similar relations had been first derived for non
stationary states, [ES], and later extended to stochastic evolutions,
[Ku], [LS]

The viewpoint of [GC] is that the reason the law holds is simply that
physical systems are chaotic to the extent that one can think that
they are Anosov systems, see [Ga1]: this is the {\it chaotic
hypothesis}; it is a version of a general principle proposed by
Ruelle, [Ru1].\annota1{\rm The hypothesis can be criticized on the
basis of various results, not only on mathematical grounds but on
physical grounds as well, see [Ga1], [RT].} Then, {\it if the
microscopic dynamics is time reversible},\annota2{\rm\ie there is an
isometry $I$ of phase space anticommuting with time evolution and with
$I^2=1$, see \equ(2.1) below.}  one can consider the f\/luctuations of
the phase space volume contraction rate\annota3{\rm Equal to minus the
logarithm $\h(x)$ of the determinant of the Jacobian matrix $\dpr
S(x)$ of the evolution map.} $\h(x)$ averaged over a time $\th$:
$$p=\fra1{\lis\h_+\,\th}\sum_{-\fra12\th}^{\fra12\th}\h(S^j x)\Eq(1.1)$$
where $\lis\h_+$ (that we suppose $>0$, see [Ru2]) is the value of the
time average of $\h(x)$ over an infinite time
$\lis\h_+=\lim_{T\to+\io}\fra1T\sum_{k=0}^{T-1}\h(S^jx)$: it is a
quantity that is $x$--independent apart from a set of points $x$ of
zero volume. We suppose, for simplicity, that the system has an
attractor that is dense on phase space (this will mean, technically,
that our systems are weakly interacting).

In the stationary state $\m$, called the SRB distribution, the
variable $p$ has f\/luctuations. Denoted $\p_\th(p)$ the probability
distribution of $p$ in the stationary state and written it, for large
$\th$, as $\p_\th(p)= e^{\lis\z(p)\th+O(1)}$, [Si], it verifies the
{\it fluctuation theorem}:
$$\fra{\lis\z(p)-\lis\z(-p)}{\lis\h_+\,p}=1,\qquad |p|<p^*\Eq(1.2)$$
for some $p^*\ge1$, see [GC]. 

The phase space contraction rate is usually identified with the {\it
entropy creation rate}, [An], [Ru2], [GC]. Therefore one sees that the
above ``law'' cannot be practically verified, for physical as well as
mathematical reasons: in fact the logarithm of the entropy creation rate
distribution $\th\, \lis\z(p)$ is, usually, not only proportional to $\th$
but also to the spatial extension of the system, \ie to the number of
degrees of freedom; so that it is extremely unlikely that observing $p$
in a large system one can see a value $p$ which is appreciably different
from $1$ (note that the normalizing constant $\lis\h_+$ in
\equ(1.1) is so chosen that the average of $p$ in the stationary state
is $1$).

In macroscopic (or just ``large'') systems the phase space contraction
rate is essentially constant (and it measures the strength of the
friction) much as the density is constant in gases at
equilibrium. Therefore one can hope to see entropy creation rate
f\/luctuations only if one can define a {\it local entropy creation rate}
$\h_{V_0}(x)$ associated with a microscopic region $V_0$ of space.

In this paper we show, heuristically, why one should expect that a
{\it local entropy creation rate can be defined and verifies a local
version of the f\/luctuation law}. We defer to \S5 the precise
definition of local entropy creation rate as its form is not important
until then.

We suppose that our system has a translation invariant spatial
structure, \eg it is a chain (or a lattice) of weakly interacting
chaotic (mixing Anosov) system. Given a finite region $V_0$ centered
at the origin and a time interval $T_0$, let $\h_+$ denote the average
density of entropy creation rate, \ie $\h_+=\lim_{V_0, T_0\to\io}
\fra1{|T_0|}\fra1{|V_0|}\sum_{j=0}^{|T_0|-1}\h_{V_0}(S^j x)$, then we set:
$$p=\fra1{\h_+ |V|} \sum_{j=-\fra12|T_0|}^{\fra12 |T_0|} \h_{V_0}(S^j
x), \qquad V=V_0\times T_0\Eq(1.3)$$
where $\h_{V_0}(x)$ denotes the entropy creation rate in the region
$V_0$, {\it a notion which we still have to define}.

Calling $\p_V(p)$ the probability distribution of $p$ in the
stationary state $\m$, \ie in the SRB distribution, {\it and assuming
that the system is a weakly coupled chain of Anosov systems} we show
that it is $\p_V(p)= e^{\z(p)|V|+O(|\dpr V|)}$ where $|\dpr V|$
denotes the size of the boundary of the space--time region $V$ and:
$$\fra{\z(p)-\z(-p)}{p\,\h_+}=1, \qquad |p|<p^*\Eq(1.4)$$
for some $\z(p)$ analytic in $p$ and for some $p^*\ge1$. It will also
result that $\lis\z(p)=r\, \z(p),\,\lis\h_+=r\, \h_+$, see \equ(1.2),
where $r$ is the ratio between the total ``volume'' of the system and
the volume $V_0$, \ie the global and local distributions are trivially
related if appropriately normalized.

The interest of the above statements lies in their independence on the
total size of the systems.

If $V_0$ is an interval of length $L=|V_0|$ and if $H=|T_0|$ then the
relative size of the error and of the leading term will be, for some
length $R$, of order $(L+H)R$ compared to order $LH$. Hence a relative
error $O(H^{-1}+L^{-1})$ is made by using simply $\z(p)$ to evaluate
the logarithm of the probability of $p$ as defined by \equ(1.3)).

{\it This means that the fluctuation theorem leads to observable
consequences if one looks at the far more probable microscopic
f\/luctuations of the local entropy creation rate} (to be yet defined:
see definition in \S5 below).

The key results for the present work are the papers [GC], [Ga3] and,
mainly, [PS]: the latter paper provides us with a deep analysis of
chains of Anosov systems and it contains, I believe, all the
ingredients necessary to make the following analysis mathematically
rigorous: however I do not attempt at a mathematical proof here.
\*

\0{\bf\S2. Reversible chains of interacting Anosov systems.}
\numsec=2\numfor=1\*

Let $(M',S')$ be a dynamical system whose phase space $M'$ is a product
of $2N+1$ identical analytic manifolds $\lis M_0$: $M'=\lis M_0^{2N+1}$
and $S': \,M' \to M'$ is a small perturbation of a product map $\lis
S_0\times \ldots\times\lis S_0\defi \tilde S_0$ on $M'$. We assume that
$(\lis M_0,\lis S_0)$ is a mixing Anosov system. The size $N$ (an
integer) will be called the ``spatial size'' of the system.

For $x,y,z\in \lis M_0$ let $F_\e(x,z,y)$ be analytic and such that
$z\to F_\e(x,z,y)$ is a map, of
$\lis M_0$ into itself, {\it $\e$--close to the identity} and $\e$--analytic
for $|\e|$ small enough. We suppose that, if $\xx=(x_{-N},\ldots,
x_N)\in M'$:
$$(S'\xx)_i= F_\e(x_{i-1},x_i,x_{i+1})\circ S_0 x_i\Eq(2.1)$$
where $x_{\pm(N+1)}$ is {\it identified} with $x_{\mp N}$ (\ie we
regard the chain as periodic); we call such a dynamical system a {\it
chain of interacting Anosov maps} coupled by nearest neighbors. It is
a special example of the class of maps considered in [PS].\annota4{\rm
In the paper [PS] it is assumed that {\it also} $\lis S_0$ (hence
$S_0$) is close to the identity, \eg within $\e$: such condition does
not seem necessary for the purposes of the present paper, hence it
will not be assumed.}

It is difficult, maybe even impossible, to construct a (non trivial)
reversible system of the above form: we therefore (see [Ga3]) consider
the system $(M,S)$ where $M=M'\times M'$ and define $S_0\defi\tilde
S_0\times (\tilde S_0)^{-1}$ and $S\defi S'\times (S')^{-1}$, called
hereafter the {\it free evolution} and the {\it interacting
evolution}, respectively. So that the system can be considered as time
reversible with a time reversal map $I(\xx,\yy)=(\yy,\xx)$.  Note that
the inverse map to \equ(2.1) does not have the same form. The map $S$
is, however, still in the class considered in [PS] because it can be
written as $S(\xx,\yy)_i=\big(S(\xx,\yy)_{i1},S(\xx,\yy)_{i2}\big)$
with:
$$\eqalign{
S(\xx,\yy)_{i1}=& F_\e(x_{i-1},x_i,x_{i+1})\circ S_0\, x_i\cr
S(\xx,\yy)_{i2}=& G_{\e,i}(\yy)\circ S_0^{-1} y_i\cr
}\Eq(2.2)$$
where $G$ has ``short range'', \ie $|G_\e(\yy)_i-G_\e(\yy')_i|$ is of
order $\e^k$ if $\yy$ and $\yy'$ coincide on the sites $j$ with
$|j-i|\le k$. By definition the system $(M,S)$ is ``{\it
reversible}'', \ie:
$$ I S=S^{-1} I,\qquad I^2=1\Eq(2.3)$$
Therefore the points of the phase space $M$ will be $(\xx,\yy)=
(x_{-N},y_{-N},\ldots, x_N,y_N)$: however, to simplify notations, we
shall denote them by $\xx=(x_{-N},\ldots,x_N)$, with $x_j$ denoting,
of course, a {\it pair} of points in $\lis M_0$.

If $\e$ is small enough the interacting system will still be
hyperbolic, \ie for every point $\xx$ it will be possible to define a
stable and an unstable manifolds $W^s_\xx,W^u_\xx$, [PS], so that the
key notion of ``Markov partition'', [Si] (see also [Ga2]), will make
sense and it will allow us to transform, following the work [PS], the
problem of studying the statistical properties of the dynamics of the
system into an equivalent, but much more familiar, problem in
equilibrium statistical mechanics of lattice spin systems interacting
with short range forces. The reduction of the dynamical nonequilibrium
problem to a short range lattice spin system equilibrium problem is
the content of what follows up to \S5, where the new application to a
local f\/luctuation theorem is presented.

Let $\lis \PP_0=(E^0_1,\ldots,E^0_{\NN_0})$ be a Markov partition, see
[Si], for the unperturbed ``single site'' system $(\lis M_0\times \lis
M_0, \lis S_0\times \lis S_0^{\,-1})$. Then
$\lis\PP_0^{2N+1}=\{E_\a\}$, $\a=(\r_{-N},\ldots,\r_N)$ with $E_\a=
E^0_{\r_{-N}}\times E^0_{\r_{-N+1}}\times\ldots \times E^0_{\r_N}$ is
a Markov partition of $(\lis M_0^{2(2N+1)}, S_0)$.

The perturbation, {\it if small enough}, will deform the partition
$\lis\PP_0^{2N+1}$ into a Markov partition $\PP$ for $(M,S)$ changing
only ``slightly'' the partition $\lis\PP_0^{2N+1}$. The work [PS]
shows that the above ``$\e$ small enough'' {\it mean that $\e$ has to
be chosen small but that it can be chosen $N$--independent}, as we
shall always suppose in what follows.

Under such circumstances we can establish a correspondence between
points of $M$ that have the same ``symbolic history'' (or ``symbolic
dynamics'') along $\lis\PP_0^{2N+1}$ under $S_0$ and along $\PP$ under
$S$; we shall denote it by $h$; see [PS].

\*
\0{\bf\S3. Operations of continuation in the symbolic dynamics.}
\numsec=3\numfor=1\*

The Markov partition $\lis\PP_0^{2N+1}$ for $S_0$ associates with each
point $\xx=(x_{-N},\ldots, x_N)$ a sequence $(\s_{i,j})$, $i\in
[-N,N], j\in (-\io,\io)$ of symbols so that $(\s_{i,j})_{j=-\io}^\io$
is the free symbolic dynamics of the point $x_i$. We call the first
label $i$ of $\s_{i,j}$ a ``space--label'' and the second a
``time--label''.  Not all sequences can arise as histories of points:
however (by the definition of $h$, see \S2) precisely the same
sequences arise as histories of points along $\PP_0$ under the free
evolution $S_0$ or along $\PP$ under the interacting evolution $S$.

The map $h$ is H\"older continuous and ``short ranged'':
$$|h(\xx)_i-h(\xx')_i|\le C\sum_j \e^{|i-j|\g'} |x_j-x'_j|^\g\Eq(3.1)$$
for some $\g,\g',C>0$, [PS], if $|x-y|$ denotes the distance in $\lis
M_0\times\lis M_0$ (\ie in the single site phase space).

Furthermore the code $\xx\otto\V\s$ associating with $\xx$ its
``history'' or ``symbolic dynamics'' $\V\s(\xx)$ along the partition
$\PP$ under the map $S$ is such that, fixed $j$:
$$\V\s(\xx)_i=\V\s(\xx')_i\ {\rm for}\  |i-j|\le\ell \qquad \tto\qquad
|x_j-x'_j| \le C\e^{\g \ell}\Eq(3.2)$$
The inverse code associating with a history $\V\s$ a point with such history
will be denoted $\xx(\V\s)$.

If $\xx=(x_{-N},\ldots, x_N)$ is coded into
$\V\s(\xx)=(\V\s_{-N},\ldots,\V\s_N)=(\s_{i,j})$, with
$i=-N,\ldots,N$, and $j\in (-\io,+\io)$, the short range property
holds also in the time direction. This means that, fixed $i_0$:
$$\s_{i,j}=\s'_{i,j}\ {\rm for}\  |i-i_0|<k, |j|<p\qquad\tto\qquad
|\xx(\V\s)_{i_0}-\xx(\V\s')_{i_0}|\le C \e^{\g k} e^
{-\k p}\Eq(3.3)$$
for some $\k,\g,C>0$, [PS] lemma 1. The constants $\k,\g,C,C',B,B'>0$
above and below should not be thought to be the same even when denoted
by the same symbol: however they could be {\it a posteriori} fixed so
that to equal symbols correspond equal values.

By construction the codes $\xx\otto\V\s(\xx)$ commute with time
evolution.

The sequences $(\s_{i,j})$ which arise as symbolic dynamics along
$\lis\PP_0$ under the free single site evolution of a point $x_i$ are
subject to constraints, that we call ``vertical'', imposing that
$T^0_{\s_{i,j},\s_{i,j+1}}\=1$ for all $j$, if $T^0_{\s,\s'}$ denotes
the ``compatibility matrix'' of the ``free single site evolution''
(\ie $T^0_{\s,\s'}=1$ if the $\lis S_0\times \lis S_0^{\,-1}$ image of
$E_\s$ intersects the interior of $E_{\s'}$ and $T^0_{\s,\s'}=0$
otherwise). We call the latter condition a ``compatibility condition''
for the spins in the $i$--th column.

The mixing property of the free evolution immediately implies that a
large enough power of the compatibility matrix $T^0$ has all entries
positive. This means that for each symbol $\s$ we can find
semiinfinite sequences:
$$\eqalign{
\s_B(\s)=&(\ldots,\s_{-1},\s_0\=\s), \qquad
T^0_{\s_{i-1},\s_i}=1,\quad {\rm for\ all}\ i\le0\cr
\s_T(\s)=&(\s\=\s_0,\s_1,\ldots), \qquad
T^0_{\s_{i},\s_{i+1}}=1,\quad {\rm for\ all}\ i\ge0\cr}\Eq(3.4)$$
and defines two functions $\s_B,\s_T$, called ``compatible
extensions'', defined on the set $\{1,\ldots,\NN_0\}$ of labels of the
single site Markov partition $\lis\PP_0$, with values in the compatible
semiinfinite sequences.

In fact there are (uncountably) many ways of performing such compatible
extensions ``from the bottom'' and ``from the top'' of the symbol $\s$
into semiinfinite compatible sequences of symbols. We imagine to select
one pair $\s_B,\s_T$ arbitrarily, once and for all, and call such a
selection a ``choice of boundary conditions'' or ``of extensions'', on
symbolic dynamics, for reasons that should become clear shortly.

We shall therefore be able to ``extend in a standard way'' any finite
compatible block\annota5{\rm A block $(\s_{i,j}),\,(i,j)\in Q$, is
naturally said to be ``compatible'' if $T^0_{\s_{i,j},\s_{i,j+1}}=1$
for all $(i,j)\in Q$ such that $(i,j+1)$ is also in $Q$.} $Q$ of
spins:
$$\V\s_Q=(\s_{i,j})_{i\in L, j\in K}, \qquad L=(a-\ell,a+\ell), \
K=(b-m,b+m)\Eq(3.5)$$
by setting $\s_{i,j}=\s_B(\s_{i,b-n})_{b-n-j}$ for $j<b-n$ and $\s_{i,j}=
\s_T(\s_{i,b+n})_{j-b-n}$ for $j>b+n$. Here $a,b,\ell,m$ are integers.

In the free evolution there are no ``horizontal'' compatibility
constraints; hence it is always possible to extend the finite block
$\V\s_Q= (\s_{i,j})_{i\in L, j\in K}$ to a ``full spin configuration''
sequence $(\s_{i,j})_{i\in [-N,N], j\in (-\io,\io)}$, obtained by
continuing the columns in the just described standard way, using the
boundary extensions $\s_B,\s_T$, above the top and below the bottom,
into a biinfinite sequence and also by extending the spin
configuration to the right and to the left to a sequence with spatial
labels running over the full spatial range $[-N,N]$. One simply
defines $\s_{i,j}$ for $i\not\in L$ as {\it any} (but prefixed once
and for all) compatible biinfinite sequence of symbols (the same for
each column).

The allowed symbolic dynamics sequences for the free dynamics (on
$\PP_0$) and for the interacting dynamics (on $\PP$) {\it coincide}
because the free and the interacting dynamics are conjugated by the
map $h$, [PS]. Therefore the above operations make sense {\it also} if
the sequences are regarded as symbolic sequences of the interacting
dynamics, as we shall do from now on.

To conclude: given a ``block'' $\V\s_Q$ of symbols, with space--time
labels $(i,j)\in Q=L\times K$, we can associate with it a point
$\xx\in M$ whose symbolic dynamics is the above described standard
extension of $\V\s_Q$. The latter depends only on the values of
$\s_{i,j}$ for $j$ at the top or at the bottom of $Q$ and, of course, on
the boundary conditions $\s_B,\s_T$ chosen to begin with.
\*
\0{\bf\S4. Expansion and contraction rates.}
\numsec=4\numfor=1\*

Consider the rates of variation of the phase space volume,
$\l_0(\xx)$, or, respectively, of the surface elements of the stable
and unstable manifolds $\l_s(\xx)$ and $\l_u(\xx)$ at the point $\xx$:
they are the logarithms of the Jacobian determinants $\dpr S(\xx)$,
$\dpr_{(\a)} S(\xx)$, $\a=s,u$, where $\dpr_{(\a)} $ denotes the
Jacobian of $S$ as a map of $W^\a_\xx$ to $W^\a_{S\xx}$ where $\a=u,s$
distinguishes the unstable manifold $W^u_\xx$ of $\xx$ or the stable
manifold $W^\s_\xx$ of $\xx$:
$$\l_\a(\xx)=-\log|\det \dpr_{(\a)} S(\xx)|,\qquad \a=0,u,s\Eq(4.1)$$
where $\dpr_{(0)} S(\xx)\defi\dpr S(\xx)$.

A hard technical problem is to represent $\l_\a(\xx)$ in terms of the
``symbolic history'' of $\xx$ along $\PP$, \ie in terms of compatible
sequences $\V\s=(\s_{i,j})$ with $i\in (-N,N),\, j\in(-\io,\io)$.
The rates $\l_\a(\xx)$ can be expressed as:
$$\l_\a(\xx)=-\log \big| \det \fra{\dpr S}{\dpr \xx}\big|_{W^\a(\xx)}=
\sum_{L\subset [-N,N]} \tilde \d_L^{(\a)}(\xx_L)\Eq(4.2)$$
where $L$ is an interval in $[-N,N]$ (with $\pm(N+1)$ identified with
$\mp N$), [PS].

For $\a=0$ this can be done by noting that the matrix $J=\fra{\dpr
S}{\dpr x}$ has an almost diagonal structure:
$J(\xx)=J_0(\xx)(1+\D(\xx))$ where $J_0(\xx)$ is the Jacobian matrix of
the free motion $J_0(\xx)=\lis J_0(x_{-N})\times \lis J_0(x_{-N+1})\times
\ldots\times \lis J_0(x_N)$ if $\xx=(x_{-N},
\ldots,x_N)$ and if $D=\big(\prod_{j=-N}^N \det \lis J_0(x_j)\big)$:
$$\det J= D \cdot
e^{{\rm Tr\,}  \log(1+\D(\xx))}=
D\cdot e^{\sum_{k=1}^\io\fra{(-1)^{k-1}}k {\rm Tr\,}\D(\xx)^k}
\Eq(4.3)$$
which leads to \equ(4.2) if one uses that the matrix elements
$\D_{p,q}=J_0^{-1}(\xx)\dpr_{x_p}\dpr_{x_q} J(\xx)$ are essentially
local, \ie bounded by $B\,(C\e)^{|p-q|\g}$ for some $\g,C,B>0$ (see
\equ(2.1),\equ(2.2), \equ(3.3)).

For $\a=u,s$ \equ(4.2) can be derived in a similar way using also
that:
\*
\0(1) the stable and unstable manifolds of $\xx$ consist of points $\yy$
which have eventually, respectively towards the future or towards the
past, the same history of $\xx$,
\\(2) they are described in a local system of coordinates around
$\xx=(\ldots,x_{-1},x_0,x_1,\ldots)$ by smooth ``short range''
functions. Suppose, in fact, that on each factor $M_0$ we introduce a
local system of coordinates $(\a,\b)$ around the point $x_i\in M_0$,
such that the unperturbed stable and unstable manifolds are described
locally by graphs $(\a,f_s(\a))$ or $(f_u(\b),\b)$.

The unperturbed stable and unstable manifolds will be smooth graphs
$(\a_i,f_s(\a_i))$ or $(f_u(\b_i),\b_i)$ with $\a_i$ varying close to
$\lis\a_i$ and $\b_i$ close to $\lis\b_i$, with $(\lis
\a_i,\lis\b_i)$ being the coordinates of $x_i$.

Fixed a point $\xx=(x_{-N},\ldots,x_N)$ with coordinates
$(\lis\a_i,\lis\b_i)_{i=-N,\ldots,N}$ the perturbed manifol\/ds of the
point $\xx$ will be described by smooth (at least $C^2$ and in fact of
any prefixed smoothness if $\e$ is sufficiently small) functions
$W^s(\V \a), W^u(\V\b)$ of $\V\a=(\a_i)_{i=-N,N}$ or of
$\V\b=(\b_i)_{i=-N,N}$ which are ``local''; \ie if $\V\a$ and $\V\a'$
agree on the sites $i-\ell,i+\ell$ or if $\V\b$ and $\V\b'$ agree on
the sites $i-\ell,i+\ell$ then:
$$\eqalign{
&||W^u(\V\b)_i-f_u(\b_i)||_{C^2}< C\e,\qquad
||W^u(\V\b)_i-W^u(\V\b')_i||_{C^2}< C\e^\ell\cr
&||W^s(\V\a)_i-f_s(\a_i)||_{C^2}< C\e,\qquad
||W^s(\V\a)_i-W^s(\V\a')_i||_{C^2}< C\e^{\ell}\cr}\Eq(4.4)$$
for some $C>0$, see [PS] lemmata 1,2. Here the norms in the first
column are the norms in $C^2$ as functions of the arguments $\V\b$ or
respectively $\V\a$, while the norms in the second column are $C^2$
norms evaluated (of course) after identifying the arguments of $\V\b$
(or $\aa$) and $\V\b'$ (or $\aa'$) with labels $j$ such that
$|i-j|\le\ell$.

\0(3) If we consider the dependence of the planes tangent to the
stable and unstable manifolds $W^s_\xx,\,W^u_\xx$ at $\xx$ we
find that they are H\"older continuous as functions of $\xx$:
$$|(d W^\a_\xx)_i-(d W^\a_\yy)_i|< C\,\sum_j
\e^{|i-j|\k}|x_j-y_j|^\g,\qquad \a=u,s\Eq(4.5)$$
where $(d W^\a_\xx)_i$ denoted the components relative to the $i$--th
coordinate of $\xx$ of the tangent plane to $W^\a_\xx$ and $C,\k,\g>0$.
\*

The above properties and the H\"older continuity \equ(3.1),
\equ(3.2), \equ(3.3) imply that the ``horizontal potentials''
$\tilde\d^{(\,\a)}_L(\xx_L)$ in \equ(4.2) are ``short ranged'':
$$|\,\tilde\d^{(\,\a)}_L(\xx_L)|\le B\,(C\e)^{(|L|-1)\g},\qquad \a=u,s
\Eq(4.6)$$
for some $B,C,\g >0$; we denote $|L|$ the number of points in the set
$L$.

We shall use the symbolic representation of $\xx\in M$ to express the
rates $\l^{(\a)}(\xx)$. For this purpose let $\xx=(x_i)_{i=-N,N}$ and
suppose that such $\xx$ corresponds to the symbolic dynamics sequence
$\V\s=(\V\s_j)_{j=-\io}^\io$ where
$\V\s_j=(\s_{-N,j},\ldots,\s_{N,j})$. We denote $\V\s_L$ the sequence
$\V\s_L=(\s_{i,j})_{i\in L,j=-\io,\io}$.

Then $\V\s_L$ {\it does not} determine $\xx_L$ (unless there is no
interaction, \ie $\e=0$): however the short range property, \equ(3.3),
of the symbolic codes and of the map $h$ conjugating the free
evolution and the interacting evolution shows that, if $L'$ is a
larger interval containing $L$ and centered around $L$, then the
sequence $\V\s_{L'}$ determines each point of $\xx_L$ within an
approximation $\le(C\e)^{(|L'|-|L|)\g}$.  Hence we can define
$\widehat\d_L^{(\,\a)}(\V\s_L)$ so that:
$$\eqalign{
\tilde\d^{(\a)}_L(\xx_L)=&\sum_{L'\supset L}
\widehat\d^{(\,\a)}_{L'}(\V\s_{L'}),\qquad
|\widehat\d_L^{(\,\a)}(\V\s_L)|< B'\,(C'\e^\g)^{|L|-1}\cr
\l_\a(\xx)=&\sum_L 2^{|L|}\widehat\d_L^{(\,\a)}(\V\s_L)\cr}
\Eq(4.7)$$
for some $B',C',\g$. This leads to expressing $\l_\a(\xx)$ in terms of
the symbolic dynamics of $\xx$ and of the ``space--localized''
potentials $\widehat \d_L^{(\,\a)}(\V\s_L)$.

Let $Q_n=L\times K$ where $K=[-n,n]$ is a ``time--interval'' and set
$$\LL^{\,\a}_{Q_n}(\V\s_{Q_n})\defi \widehat
\d_L^{(\,\a)}([\V\s_{Q_n}])-\widehat \d_L^{(\,\a)}([\V\s_{Q_{n-1}}])\Eq(4.8)$$
if $n\ge1$ and $[\V\s_{Q_n}]$ denotes a {\it standard extension} (in the
sense of \S3) of $\V\s_{Q_n}$; or just set $\LL^{\,\a}_{Q_0}\defi \widehat
\d_L^{(\,\a)}([\V\s_{Q_0}])$ for $n=0$. We define
$\LL^\a_Q(\V\s_{Q})$ for $Q=L\times K$ and $K$ not centered (\ie
$K=(a-n,a+n),\, a\ne0)$ so that it is translation invariant with
respect to space time translations (of course the horizontal
translation invariance is already implied by the above definitions and
the corresponding translation invariance of
$\widetilde\d^{(\,\a)}_L$).

The {\it remarkable property}, consequence of the H\"older continuity
of the functions in \equ(4.1) and of the \equ(3.3),\equ(4.7), see
[PS], is that for some $\g,\k,B,C>0$:
$$|\LL^\a_Q(\V\s_Q)|\le B\,( C \e^\g)^i\, e^{-\k j}\Eq(4.9)$$
if $i,j$ are the horizontal and vertical dimensions of $Q$.

In this way we define a ``space--time local potential'' $\LL^{(\a)}_Q$
which is, by construction, translation invariant and such that, if
$\L$ denotes the box $\L=[-N,N]\times[-M,M]$ the following
representations for the rates in \equ(4.1) hold:
$$-\log|\det\dpr_{(\,\a)}{S^{2M+1}}(S^{-M}\xx)|=\sum_{Q\subset \L}
\LL^\a_Q(\V\s_Q)+ O(|\dpr\L|)\Eq(4.10)$$
where $O(|\dpr\L|)$ is a ``boundary correction'' due to the fact that in
\equ(4.10) one should really extend the sum over the $Q$'s centered at
height $\le M$ and contained in the infinite strip
$[-N,N]\times[-\io,\io]$ rather than restricting $Q$ to the region
$\L$. Hence the remainder in \equ(4.10) can, in principle, be
explicitly written, in terms of the potentials $\LL_Q^{(\,\a)}$, in the
boundary term form usual in Statistical Mechanics of the
$2$--dimensional short range Ising model and it can be estimated to be
of $O(|\dpr\L|)$ by using \equ(4.9).
\ifnum\mgnf=1\pagina\fi
\*

\0{\bf\S5. Symmetries. Local entropy creation. SRB states and f\/luctuations.}
\numsec=5\numfor=1\*

Besides the obvious translation invariance symmetry the dynamical
system has a {\it time reversal symmetry}; this is the diffeomorphism
$I$, see \equ(2.3), which {\it anticommutes} with $S$ and $S_0$:
$$I S=S^{-1} I,\qquad I S_0=S_0 I^{-1},\qquad I^2=1\Eq(5.1)$$
We can suppose that the Markov partition is time reversible, \ie to each
element $E_{\V\s}$ of the partition $\PP$ one can associate an element
$E_{\V \s'}=I E_{\V \s}$ which is {\it also} an element of the
partition. Here we simply use the invariance of the Markov partition
property under maps that either commute or anticommute with the
evolution $S$: hence it is not restrictive, see [Ga2],[Ga3], to suppose
that for each $\V\s$ one can define a $\V\s'$ so that $E_{\V\s'}=I
E_{\V\s}$. We shall denote such $\V\s'$ as $I\V\s$ or also $-\V\s$. For
$\e=0$, \ie for vanishing perturbation, the map $I$ will act
independently on each column of spins of $\V\s$. This property remains
valid for small perturbations; hence:
$$I\V\s=\{\s'_{i,j}\}=\{-\s_{i,-j}\}\defi -\V\s^I\Eq(5.2)$$
\ie time reversal simply ref\/lects the spin configuration corresponding
to a phase space point and changes ``sign'' of each spin.

The functions $\l_\a(\xx)$ and their ``potentials'' $\LL^\a_Q(\V\s_Q)$
verify, as a consequence, if $Q=[-\ell,\ell]\times[-k,k]$ is a
centered rectangle:
$$\l_{\a}(I\xx)=-\l_{\a'}(\xx),\qquad
\LL^\a_{Q}(\V\s_Q)=-\LL^{\a'}_Q(-\V\s_Q^I)\Eq(5.3)$$
where $\a'=s$ if $\a=u$ and $\a'=u$ if $\a=s$, $\a'=0$ if $\a=0$.  The
above symmetries will be translated into remarkable properties of
the SRB distribution.

\*
\0{\it Definition:
Fixed a point $\xx=(\ldots, x_{\ell-1}, x_{\ell}, x_{\ell+1},\ldots)$
consider the map \equ(2.1) as a map of
$\xx_{V_0}\defi(x_j)_{j\in V_0}=(x_{-\ell},\ldots,x_\ell)$ into:
$$x'_{V_0}=S(\ldots,x_{-\ell-1},\xx_{V_0},x_{\ell+1},\ldots)_{V_0}
\Eq(5.4)$$
defined by \equ(2.1) for $i\in[-\ell,\ell]$. We call {\it ``local
entropy production rate''} associated with the ``space like box''
$V_0=[-\ell,\ell]$ at the phase space point
$\xx=(\ldots,x_{\ell-1},x_{\ell},x_{\ell+1},\ldots)$ the quantity
$\h^0_{V_0}(\xx)$ equal to {\sl minus the logarithm of the determinant
of the $2(2\ell+1)\times2(2\ell+1)$ Jacobian matrix of the map},}
\equ(5.4).
\*

Likewise we can consider the corresponding Jacobian determinants of
the restriction of the map $S$ to the stable and unstable manifolds of
$\xx$. Such determinants will depend not only from $x_i$, $i\in V_0$, and
on the nearest neighbors variables $x_{\pm\ell}$ but {\it also} on the
other ones $x_k$ with $|k|>\ell+1$: however their dependence from the
variables with labels $|k|>\ell$ is exponentially damped as
$\e^{(|k|-\ell)\g}$, by \equ(4.9).

If we look at the average phase space variation rates
$\h^0_{V_0},\h^s_{V_0},\h^u_{V_0}$ between the time $-\th$ and $\th$ we
can find, via a power expansion like the one in \equ(4.3) along the
lines leading from \equ(4.3) to \equ(4.10), a mathematical expression as:
$$\h^\a_{V_0}(\xx)\simeq
\sum_{Q}{}^*  \LL_Q^\a(\V\s_Q)\Eq(5.5)$$
where the $\sum^*_QQ$ runs over rectangles $Q$ centered at $0$--time
$Q=[a-\ell,a+\ell]\times[-k,k]$ with $[a-\ell,a+\ell]\subseteq V_0$.
This could be taken as an alternative {\it definition} of
$\h^\a_{V_0}$, as it is a rather natural expression. For our purposes,
if $V=V_0\times[-\th,\th]$, one needs to note that \equ(5.5) holds at
least in the sense that:
$$\fra1{V_0\cdot(2\th+1)}\sum_{j=-\th}^\th \h^{(\,\a)}_{V_0}(S^j\xx)=
\fra1{V_0\cdot(2\th+1)}\sum_{Q\subset V} \LL^\a_Q(\V\s_Q)+ \fra{O(|\dpr
V|)}{|V|}\Eq(5.6)$$
\ie expression \equ(5.5) can be used to compute the average local entropy
creation rate in the space--time region $V$ {\it up to boundary
corrections $O(|\dpr V|)$} (that can be neglected for the purposes of
the following discussion).

We now study the SRB distribution $\m$: denoting by $\media{F}_+$ the
average value with respect to $\m$ of the observable $F$ we can say,
see [Si], [PS], that if $\L=[-N,N]\times[-T,T]$:
$$\media{F}_+=\lim_{T\to\io} \fra{\sum_{\V\s} F(\V\s) e^{\sum_{Q\subset
\L} \LL^u_{Q}(\V\s_Q)}}{\sum_{\V\s} e^{\sum_{Q\subset
\L}\LL^u_{Q}(\V\s_Q)}}\Eq(5.7)$$
We want to study the properties of the f\/luctuations of:
$$p=\fra1{V \h_+}\sum_{Q\subset V}\LL^u_{Q}(\V\s_Q), \qquad
{\rm if\ \ }\h_+=\lim_{V\to\io}\fra1V\sum_{Q\subset V}\media{\LL^u_{Q}}_+
\Eq(5.8)$$
for which we expect a distribution of the form $\p_V(p)=
\,const\,e^{V \z(p)+O(\dpr V)}$. The SRB distribution gives
to the event that $p$ is in the interval $dp$ the probability $\p_V(p)
dp$ with:

$$\p_V(p)=\,const\, \sum_{at\ fixed\ p} e^{\sum_{Q\subset \L}
\LL^u_Q(\V\s_Q)}\Eq(5.9)$$
and (defining implicitly $U^u$):
$$\eqalign{
&\sum_{Q\subset\L} \LL^u_Q(\V\s_Q)=
\sum_{Q\subset V} \LL^u_Q(\V\s_Q)+
\sum_{Q\subset \L/V}\LL^u_Q(\V\s_Q)+\,O(|\dpr V|\,\k^{-1})\defi\cr
&\defi
U_V^u(\V\s_V)+U^u_{\L/V}(\V\s_{\L/V})+O(|\dpr V|\,\k^{-1})\cr}\Eq(5.10)$$
with $\k>0$, having used the ``short range'' properties \equ(4.9) of
the potential.

In the sums in \equ(5.7) we would like to sum over $\V\s_V$ and over
$\V\s_{\L/V}$ as if such spins were independent labels. This is not
possible because of the vertical compatibility constraints. However
the mixing property supposed on the free evolution implies that the
compatibility matrix $T^0$ raised to a large power $R$ has positive
entries. Hence if we leave a gap of width $R$ above and below $V$ we
can regard as independent labels the labels $\s_{i,j}$ with $i$ in the
space part $V_0$ of the region $V=V_0\times[-\th,\th]$ and
with $|j|>\th+R$, by a distance $\ge R$ above or below the region
$V$. Denoted $V+R\defi V_0\times [-\th-R,\th+R]$ remark that:
$$\sum_{Q\subset\L} \LL^u_Q(\V\s_Q)=
U_V^u(\V\s_V)+U^u_{\L/(V+R)}(\V\s_{\L/(V+R)})+
O(|\dpr V|\,(R+\k^{-1}))\Eq(5.11)$$
Hence, proceeding as in [GC1], we change the sum over (the dummy
label) $\V\s$ in the denominator to a sum over $-\V\s^I$ and using
$\LL^u_{Q^I}(-\V\s_Q^I)= -\LL^s_{Q}(\V\s_Q)$:
$$\fra{\p_V(p)}{\p_V(-p)}= \fra{\sum_{at\ fixed\ p} e^{\sum_{Q\subset
V}\LL^u_Q(\V\s_Q)} e^{U^u_{\L/(V+R)}(\V\s_{\L/(V+R)})}}
{\sum_{at\ fixed\ p} e^{\sum_{Q\subset
V}-\LL^s_Q(\V\s_Q)} e^{U^u_{\L/(V+R)}((-\V\s^I)_{\L/(V+R)})}}
\,e^{O(|\dpr V|)}\Eq(5.12)$$
with summation being over spin configurations in the ``whole
space--time'' $\L$, subject to the specified constraint of having the
same value for $p$, \ie same average local entropy creation rate
in the space--time region $V$. The latter expression becomes, since
labels $\V\s,-\V\s^I$ (resp. in numerator and
denominator of \equ(5.12)) are independent {\it dummy labels}:
$$\fra{\sum_{at\ fixed\ p} e^{\sum_{Q\subset
V}\LL^u_Q(\V\s_Q)} Z(\L/(V+R))}{\sum_{at\ fixed\ p} e^{\sum_{Q\subset
V}-\LL^s_Q(\V\s_Q)} Z(\L/(V+R))}\,e^{O(|\dpr V|)}\Eq(5.13)$$
so that by the \equ(5.6), \equ(5.8) and since the symmetry relations
above imply $\sum_{Q\subset V}(\LL^u_Q(\V\s_Q)$ $+\LL^s_Q(\V\s_Q))=
V\,\h_+\, p$, up to corrections of size $O(|\dpr V| \k^{-1})$ we find,
(note the repetition of the comparison argument given in [GC]):
$$\fra{\p_V(p)}{\p_V(-p)}=e^{\h_+\, V\, p}\ e^{O(|\dpr V|)}\Eq(5.14)$$
yielding a {\it local fluctuation law}.
\*
\0{\bf Acknowledgments:\it\ I have profited of stimulating discussions
with F. Perroni, who also helped with numerical tests of the
above ideas, with F. Bonetto and D. Ruelle. This work is part of the
research program of the European Network on: ``Stability and
Universality in Classical Mechanics", \# ERBCHRXCT940460; partially
supported also by Rutgers University and CNR-GNFM.}
\*

\0{\bf References}
\*\parindent=0pt\parskip1mm
[An] Andrej, L.: {\it The rate of entropy change in
non--Hamiltonian systems}, Physics Letters, {\bf 111A}, 45--46, 1982.

[ES] Evans, Searles, : {\it Equilibrium microstates which generate
second law violating steady states}, Physical Review E, {\bf 50E},
1645--1648, 1994.

[Ga1] Gallavotti, G.: {\it Chaotic dynamics, f\/luctuations,
nonequilibrium ensembles}, Chaos, {\bf8}, 384--392, 1998.

[Ga2] Gallavotti, G.: {\it New methods in nonequilibrium
gases and f\/luids}, Proceedings of the conference {\sl Let's face chaos
through nonlinear dynamics}, U. of Maribor, 24 June-- 5 July 1996,
ed. M. Robnik, in print in Open Systems and Information Dynamics,
Vol. 5, 1998 (also in mp$\_$arc@ math. utexas.edu \#96-533 and
chao-dyn \#9610018).

[Ga3] Gallavotti, G.: {\it Reversible Anosov maps and large
deviations}, Mathematical Phy\-sics Electronic Journal, MPEJ, (http://
mpej.unige.ch), {\bf 1}, 12 pp., 1995.

[GC1] Gallavotti, G., Cohen, E.G.D: {\it Dynamical ensembles in
nonequilibrium statistical mechanics}, Physical Review Letters,
{\bf74}, 2694--2697, 1995. And {\it Dynamical ensembles in stationary
states}, Journal of Statistical Physics, {\bf 80}, 931--970, 1995.

[Ku] Kurchan, J.: {\it Fluctuation theorem for stochastic dynamics},
Journal of Physics A, {\bf 31}, 3719--3729, 1998.

[LS] Lebowitz, J.L., Spohn, H.: {\it The Gallavotti-Cohen Fluctuation
Theorem for Stochastic Dynamics}, preprint, Rutgers University, June
1998.

[MR] Morriss, G.P., Rondoni, L.: {\it Applications of periodic orbit
theory to $N$--particle systems}, Journal of Statistical Physics, {\bf
86}, 991, 1997.

[PS] Pesin, Y.B., Sinai, Y.G.: {\it Space--time chaos in chains of
weakly inteacting hyperbolic mappimgs}, Advances in Soviet
Mathematics, {\bf3}, 165--198, 1991.

[RT] Rom-Kedar, V., Turaev, D.: {\it Big islands in dispersing
billiard--like potentials}, pre\-print, Weizmann Institute, April 2,
1998.

[Ru1] Ruelle, D.: {\it Sensitive dependence on initial conditions
and turbulent behavior of dynamical systems}, Annals of the New York
Academy of Sciences, {\bf356}, 408--416, 1978.

[Ru2] Ruelle, D.: {\it Positivity of entropy production in
nonequilibrium statistical mechanics}, Journal of Statistical Physics,
{\bf 85}, 1--25, 1996. And: {\it Entropy production in nonequilibrium
statistical mechanics}, Comm. in Mathematical Physics,
{\bf189}, 365--371, 1997.

[Si] Sinai, Y.: {\sl Lectures in ergodic theory}, Lecture notes
in Mathematics, Prin\-ce\-ton U. Press, Princeton, 1977.
\ciao